\documentclass{sigchi}

 \toappear{
 }

\usepackage{balance}  
\usepackage{graphics} 
\usepackage[T1]{fontenc}
\usepackage{txfonts}
\usepackage{mathptmx}
\usepackage[pdftex]{hyperref}
\usepackage{color}
\usepackage{booktabs}
\usepackage{textcomp}
\usepackage{microtype} 
\usepackage{ccicons}  

\usepackage{todonotes}

\def\plaintitle{Robotic Haptic Proxies for Collaborative Virtual Reality}

\def\emptyauthor{}
\def\plainkeywords{
	Tangible User Interface;
	Physical presence;
	Virtual Reality;
	Remote collaboration;
    Haptics;
   }

\makeatletter
\def\url@leostyle{%
  \@ifundefined{selectfont}{
    \def\UrlFont{\sf}
  }{
    \def\UrlFont{\small\bf\ttfamily}
  }}
\makeatother
\def\pprw{8.5in}
\def\pprh{11in}

\definecolor{linkColor}{RGB}{6,125,233}
\hypersetup{%
  pdftitle={\plaintitle},
  pdfauthor={\emptyauthor},
  pdfkeywords={\plainkeywords},
  bookmarksnumbered,
  pdfstartview={FitH},
  colorlinks,
  citecolor=black,
  filecolor=black,
  linkcolor=black,
  urlcolor=linkColor,
  breaklinks=true,
}

\begin{document}

\title{\plaintitle}
\numberofauthors{4}
    \author{
      \alignauthor Zhenyi He\\     
      \affaddr{Future Reality Lab, New York University }  \\
      \email{zh719@nyu.edu}
      \alignauthor Fengyuan Zhu\\   
      \affaddr{Future Reality Lab, New York University }  \\
      \email{fz567@nyu.edu}
      \alignauthor Aaron Gaudette\\   
      \affaddr{Future Reality Lab, New York University }  \\
      \email{ag4678@nyu.edu}
\and
	\alignauthor Ken Perlin\\
    \affaddr{Future Reality Lab, New York University }  \\
    \email{ken.perlin@gmail.com}      
%
}
\maketitle

\begin{abstract}

We propose a new approach for interaction in Virtual Reality (VR) using mobile robots as proxies for haptic feedback. This approach allows VR users to have the experience of sharing and manipulating tangible physical objects with remote collaborators. Because participants do not directly observe the robotic proxies, the mapping between them and the virtual objects is not required to be direct. In this paper, we describe our implementation, various scenarios for interaction, and a preliminary user study.
\end{abstract}

\category{H.5.1.}{Multimedia Information Systems}{Artificial, augmented, and virtual realities} 


\section{Introduction}

In the past decade, virtual reality (VR) has reached a greater level of popularity and familiarity than ever before. VR is gaining traction in the consumer market, and integrating physical and virtual environments seamlessly is a well-recognized challenge. Haptic feedback can be a powerful component of immersion and can be used to improve user experience in VR to great effect.

In this paper, we present a generic, extensible system for providing haptic feedback to multiple users in virtual reality by intelligently directing one or more mobile robots. These robots act as proxies for real, physical objects or forces, assisting the environment by providing a deeper level of immersion at a lower cost. Using the Holojam Software Development Kit (SDK) along with OptiTrack tracking technology, we are able to determine the absolute position and rotation of various real objects in the environment, such as a user's wrists or a workstation/table. With this knowledge of global state, we utilize an outside-in approach to control lightweight robots (proxies) in realtime with a high degree of accuracy. The system is physically concise and simple to set up.

We implemented multiple prototype applications that address some of the many possibilities this technology enables. Because the Holojam SDK primarily supports multi-user VR experiences, we chose to focus on collaborative, shared experiences. We implemented prototypes involving users in the same room (shared-space) as well as prototypes involving users in different physical locations (remote-space), with support for varying spatial configurations.

We defined three possible mappings between the representation of physical proxies, which are "invisible" when wearing a VR headset, and virtual objects: one-to-one, many-to-one, and one-to-many. Multiple virtual objects may be represented physically by a single robot, or vice versa, depending on the number and distribution of the robots in the scene, as well as user proximity (shared-space versus remote-space).

In remote-space, our system allows multiple users to "share" touch on the same virtual object, enabling new forms of collaboration for remote teamwork. In shared-space, one proxy or several distributed proxies can be synthesized via gesture recognition and prediction to facilitate swift haptic response for a large number of virtual objects, as well as "commanding" of physical objects to move without touching them. We executed all of these ideas as prototype applications.

Our main contributions:
\begin{itemize}
\item Remote synchronizable robotic proxies for VR. Based on robotic assistants, people could do physical collaboration remotely.
\item Mappings between virtual objects and invisible physical robots. We extend the possibility via implementing different mapping styles. Therefore, we could control one virtual object by multiple robots, or control multiple virtual objects by only one robot.
\item Augment the experience by combination of physical feedback and virtual scene.
\end{itemize}


\section{Related Work}

Several methods have been advanced that simulate the sense of touch. However, many are either not as mobile or not as lightweight as our system.





\subsection{Haptic Interfaces}
Recently more and more research has arisen focused on the intersection between haptic feedback and collaboration. InForm \cite{follmer2013inform} proposed utilizing shape displays in multiple ways to manipulate by actuating physical objects.
Tangible Bits was proposed to empower collaboration by manipulating physical objects at 1997\cite{ishii1997tangible}, and the idea was extended in 2008\cite{ishii2008tangible}.
PSyBench, a physical shared workspace, presents a new approach to enhance remote collaboration and communication, based on the idea of Tangible Interfaces at 1998\cite{brave1998tangible}.
The concept of synchronized distributed physical objects was mentioned in PSyBench\cite{brave1998tangible}, which demonstrated the potential of physical remote collaboration. One contribution in this paper is to show how people can experience consistent physical feedback over distance, regardless of the physical configuration of the corresponding remote space.
PSyBench\cite{pedersen2011tangible} only had 1-to-1 mapping while we extended the mapping style and kind of scenarios. Also objects could not be lifted and displayed the same movement without VR support. InForm\cite{follmer2013inform} did not support collaboration and the materials are fixed on the table, while we offer a more lightweight approach. SnakeCharmer\cite{araujo2016snake} had similar ideas about one-to-many mapping. However, we support collaboration and wireless.

\subsection{Haptic Feedback}

Haptic feedback has been mentioned frequently in VR training, especially in the medical field. The sense of touch is the earliest developed in human embryology and is believed to be essential for practice \cite{fager2004use,dunkin2007surgical}. Robot-assisted minimally invasive surgery (RMIS) holds great promise for improving the accuracy and dexterity of a surgeon \cite{okamura2009haptic}. Haptic feedback has potential benefits not only in training, but in other interactions as well. Reality based interaction was proposed for post-WIMP interfaces \cite{jacob2008reality}. Tangible interactions are observable with both visual and haptic modalities that could help people utilize basic knowledge about the behavior of our physical world \cite{ullmer2005token+}.

\subsection{Haptic Re-Targeting}
Manipulating multiple virtual objects is always a challenge, in that precisely-located haptic proxy objects are required. \cite{azmandian2016haptic} proposed multiple approaches to align physical and virtual objects. Redirected touching \cite{kohli2010redirected} considered the inflexibility of passive haptic displays and introduced a deliberate inconsistency between real hands and virtual hands. In redirected touching, a single real object could provide haptic feedback for virtual objects of various shapes to enrich the mapping between virtual objects and physical proxies.

\subsection{Telepresence}
C-Slate presented a new vision-based system, which combined bimanual and tangible interaction and the sharing of remote gestures and physical objects as a new approach to remote collaboration \cite{izadi2007c}. 
\cite{he2014hand} tried an augmented reality way to control distant objects without feedback.
Also with shared workspaces that can capture and remotely render the shapes of people and objects, users can experience haptic feedback based on shaped displays \cite{leithinger2014physical}.

\section{MOBILE PHYSICAL PROXIES}
To support the illusion of physical sharing between collaborators, our system requires one or more mobile robots, which must be intelligently controlled in a manner that respects the constraints of the robots and the environment. We accomplish this with an outside-in approach, managing global state and sending robot control messages from a single server machine.

Using the ad-hoc server model in the Holojam SDK, we first ascertain global state by reading realtime tracking data from the OptiTrack system, then calculate the target positions for any robots that are currently active in the environment, based on the application requirements. Our ad-hoc server maintains as many rooms as is necessary for remote-space applications (all of the examples in this paper use either one or two rooms) and commands the connected robots to move when needed. Because both the robots and optitrack cameras are small, and the ad-hoc server can run on a midrange computer, the entire setup is fairly lightweight and portable.

\begin{figure}[h!]
  \centering
  \includegraphics[width=0.5\columnwidth]{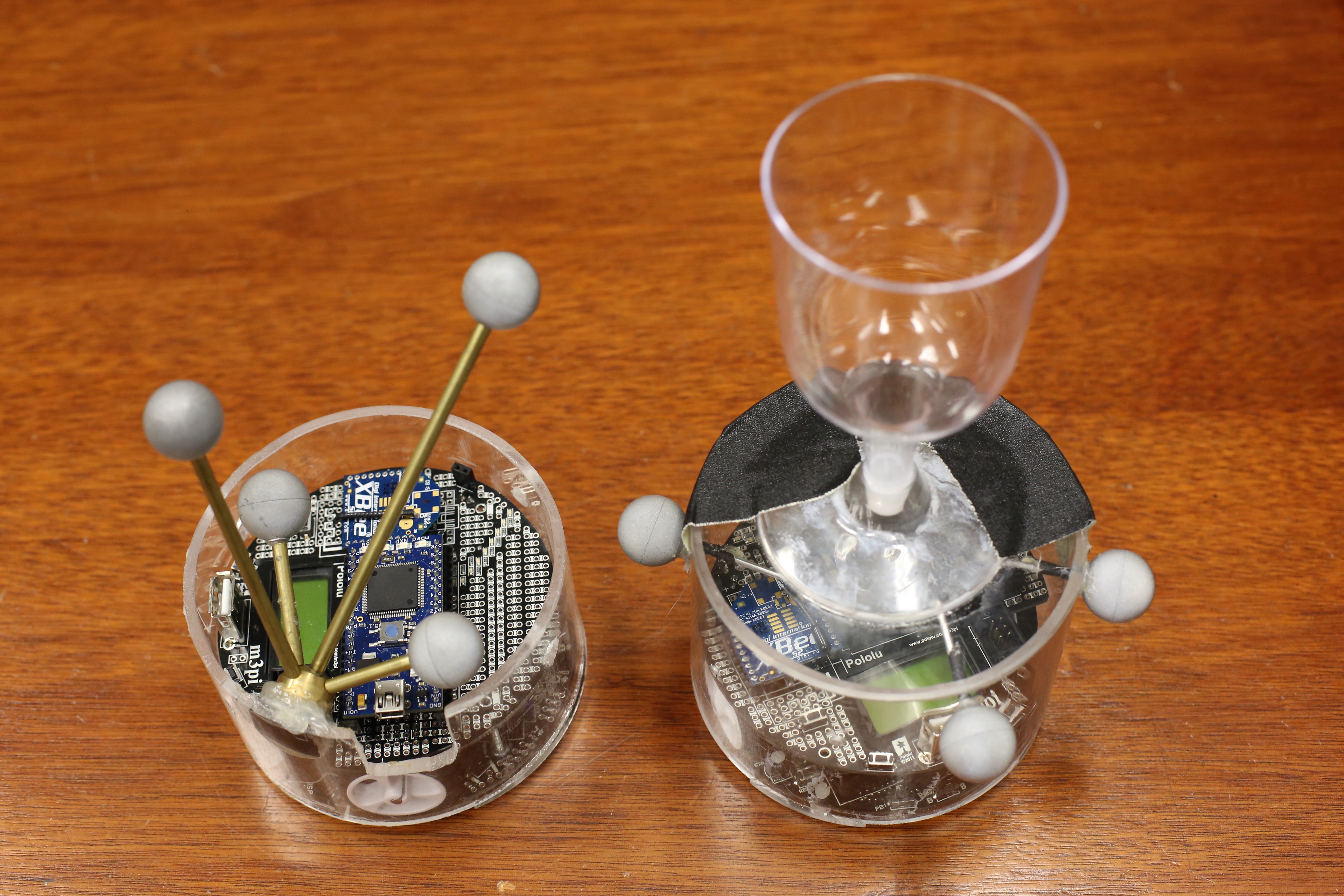}
  \caption{Robots with tracked configuration}~\label{fig:robots}
\end{figure}

\subsection{Robot Design}

Beginning with several m3pi robots, we attached a configuration of tracked markers to each one in order to determine its absolute position and orientation via the OptiTrack system. We created a cylinder-shaped acrylic cover for the robots (Figure \ref{fig:robots}) in order to facilitate easy gripping. This grip was standardized across all the robots, with a minimalist, scale-matched virtual representation available for when interaction was necessary.

\subsection{Mappings Between Physical and Virtual Space}
One challenge of physically representing a virtual environment  is the method used to map virtual objects to their physical counterparts (or vice versa). Conventionally, virtual objects have no physical representation and can only be controlled indirectly or through the use of a standard input device (such as a game controller). Also, the set of physically controlled objects that are visually represented in VR have been traditionally limited to the input devices themselves. In contrast, our system provides three different mapping mechanisms to augment the relationship between physical and virtual representation.

\begin{itemize}

\item \textbf{One-to-One Mapping} \\
This is the standard and most straightforward option. When users interact with a virtual object in the scene, they simultaneously interact with a physical proxy at the same location.

\item \textbf{Many-to-One Mapping} \\
When multiple proxies represent one virtual object, we define the mapping as "many-to-one." This is useful for remote-space applications: a virtual object could exist in the shared environment, which could then be manipulated by multiple users via their local proxies.

\item \textbf{One-to-Many Mapping} \\
Various constraints including cost and space limit the capability of maintaining a physical counterpart for every one of a large number of virtual objects. When less proxies are available than virtual objects, one of the total available proxies could represent a given virtual object when required. For example, a user with a disordered desk in VR may want to reorganize all of their virtual items. In each instance of the user reaching to interact with a virtual item, the single (or nearest, if there are multiple) proxy would relocate to the position of the target item, standing by for pickup or additional interactions. The virtual experience is seamless, while physically a small number of proxies is constantly in motion.

\end{itemize}

\subsection{Localization of User Environments}
Our system enables teams to collaborate with each other in remote-space. Here we assume the environment is a workstation for the sake of example. One issue is that not all users may share the same spatial configuration. To address this, we built a generic model for synchronizing proxies on a table.


\begin{figure}[h!]
  \centering
  \includegraphics[width=0.95\columnwidth]{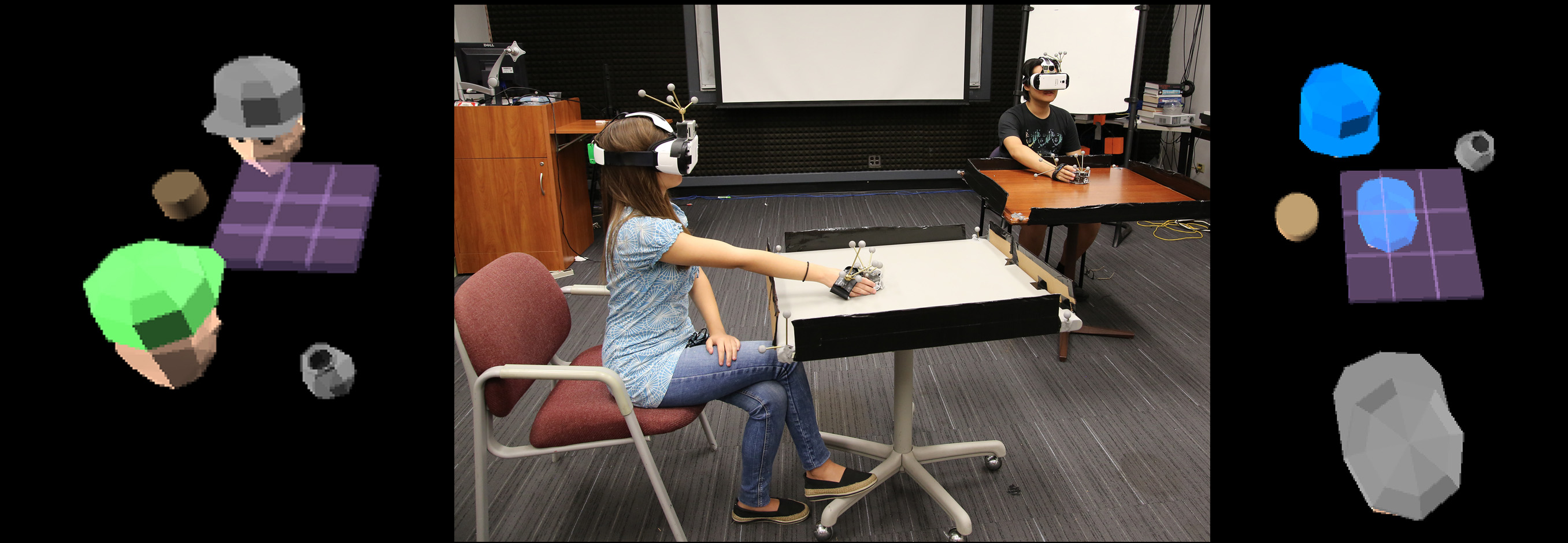}
  \caption{Simulated remote-space setup with both VR views}~\label{fig:vrconfig}
\end{figure}

In a demonstration, we assigned two participants to seats arranged perpendicular to each other, with tables unequal in size (Figure 2). In VR, each participant observed the other as directly opposite themselves, with a consistent table representation. This design allows users to experience a natural viewpoint regardless of physical setup. The robotic proxies, meanwhile, adjust themselves to operate within the minimum boundary, relative to each user's individual view.

We also chose to introduce latency for synchronizing user actions in remote-space. Movements involving the proxy are delayed by a specified time interval (we use a delay of one second for the demonstration above), to give the robot sufficient time to catch up with each user's actions. This smooths the haptic operations considerably.

We added "artificial" latency to improve the user experience. Based on the robot we chose and the size of workstation, our latency test concluded ranges from 1-1.5 seconds. Therefore we "faked" the scene that user supposed to see in order to eliminate the feeling of true latency. 

In the experiment section we test users on their perception of this latency, with positive results. For shared-space applications, we forego latency and instead render a "loading" sprite to signify to the user that the proxy has not yet arrived at its final destination.


\section{Prototype Applications}

\subsection{Pass the Mug}
In this scenario, we demonstrate a novel method to extend the capability of human interaction with physical and virtual objects. We employ a custom gesture recognition technique alongside our physical proxies, enabling users to command the position of a mug using hand motions. Utilizing a simple configuration of tracked markers that reports each user's wrist position and orientation via the OptiTrack system, users can push a proximate mug across the table via a pushing gesture, pull a distant mug towards them with a pulling gesture, or motion in another direction to slide the mug towards another part of the table. Users must first select a target by aiming their palm at an object--the object will shake to indicate it is being targeted--afterwards the object will glow to indicate it is under gesture mode. Users can still pick up objects and interact with them normally when they are close enough to reach.

\begin{figure}
  \centering
  \includegraphics[width=0.9\columnwidth]{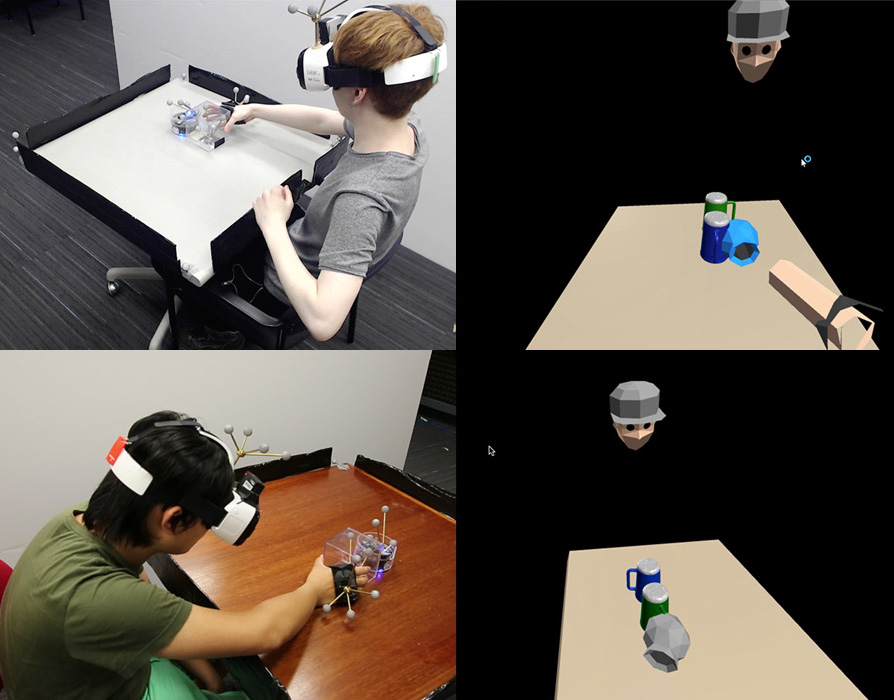}
  \caption{"Clinking Drinks" Scenario}~\label{fig:cheers}
\end{figure}

\subsection{Clinking Drinks}
This scenario utilizes a one-to-one mapping, simulating the effect of two users striking their mugs together while in separate physical locations (see in Figure \ref{fig:cheers}). In each local environment, the user's mug's position is governed by a non-robotic tracked physical object, and the force of the strike is simulated via a synchronized moving proxy.

\subsection{Tic-Tac-Toe}
\begin{figure}[h!]
  \centering
  \includegraphics[width=1\columnwidth]{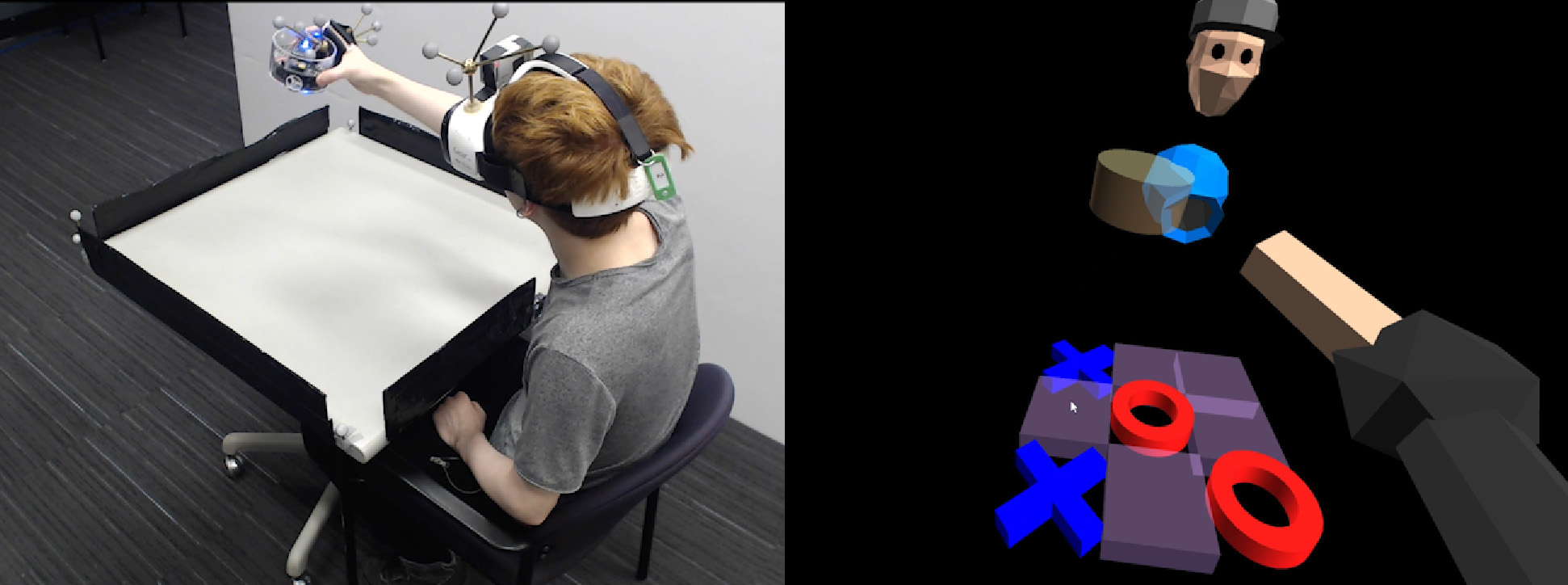}
  \caption{"Tic-Tac-Toe" Scenario}~\label{fig:ttt}
\end{figure}

Here we recreated the classic board game, \textit{Tic-Tac-Toe}, in VR, but for users in remote locations. We utilize a "controller" object, allowing players to select a tile for each game step (see the transparent cylinder holding by user in Figure \ref{fig:ttt}). This application implements many-to-one mapping: two proxies are synchronized all the time, with a single virtual controller synchronized across both spaces so that a player can reach for it on their turn and experience haptic feedback. When one player is holding the controller and considering their next move, the other player can observe this movement in their view as well. Each game step, a player's selected tile displays either an "O" or an "X" shape. Turns are processed sequentially. The techniques used in this prototype could easily be expanded to other tabletop games.

\subsection{City Builder}
The City Builder demo visualizes a virtual 3D miniature of a city on a flat surface, where users are able to pick up any building and move it to somewhere else on the map. Each user's wrist is tracked in the same manner as in the "Pass the Mug" demo, and a proxy is fitted behind-the-scenes to the virtual building which is visually nearest to a given user's hand -- an example of one-to-many mapping. In this implementation, the movement and position of the robots is completely invisible to the user.



\section{Experiment and Results}
We designed two tests for our user study, modified from our prototype applications. The first, a duplicate of our "Tic-Tac-Toe" application, demonstrates many-to-one mapping. We included a version of \textit{Tic-Tac-Toe} that was purely virtual (no haptics) for comparison. The second experiment, "Telekinesis", was loosely based on our "Pass the Drinking Mug" application, demonstrating one-to-one mapping, and measuring users' ability to learn the gesture system and to command a target at various locations.

Following each test, users filled out a survey of their experience. Of the twelve questions on the survey, three were general, six regarded the \textit{Tic-Tac-Toe} test, and three regarded the Telekinesis test.

\subsection{Hardware and Software}
Participants wore Samsung GearVR headsets with OptiTrack tracked markers, as per the current specifications of the Holojam SDK. 
Holojam\cite{perlin2016future} is an untethered virtual reality headset system that enables people have shared-space VR experience. Holojam has been demoed in SIGGRAPH 2015 performance session. Briefly: each participant needs to put on a lightweight wireless motion-tracked GearVR headset as well as strap-on glove markers, tracked by OptiTrack. These devices allow them to see everyone else as an avatar, walk around the physical world, and interact with real physical objects. You can see the headset and paired glove in Figure \ref{fig:vrconfig}. 

We can calculate the head's position based on marker information. The users are tracked all the time. In the framework of Holojam, we have a server to receive data from outside like OptiTrack and send customized message to all clients (phones). That is why users wearing headset could see real-time scene. The markers we used are only fixed on headsets and hand-made glove. Users could only feel the glove itself instead of the markers, see Figure \ref{fig:vrconfig}. Actually it is just a band around the palm. From the feedback, no users complained on the simplified glove.

In our design, we need a lightweight physical object, which is both tracked and movable. The m3pi robot\cite{Pololu} is a complete, high-performance mobile platform with an m3pi expansion board as its second level. You can see the combination of the m3pi robots and markers in Figure \ref{fig:robots}.

For both experiments, the participants' wrists were tracked by hand straps containing tracked markers, and two tables of different sizes were also tracked.

\subsection{Participants}
Sixteen participants invited to join our study, three female. Ages ranged from twenty-two to fifty-two, and the median age was twenty-six years old. All participants were right-handed. Seventy-five percent had experienced VR prior to the study.

\subsection{Tic-Tac-Toe}

In this experiment, the participant experienced a multiplayer (two-person) playthrough of both our prototype application version of \textit{Tic-Tac-Toe} and a purely virtual version, where their tracked hand was used for tile selections instead of the controller.

The time elapsed for each game was recorded, as well as the participant's responses to the following questions:
\begin{itemize}
\item Did you feel like you and your testmate were sitting at different tables?
\item Did you feel like your opponent was moving naturally?
\item Were you more comfortable with the physical or the virtual version of the game?
\end{itemize}

On a scale from one to five:
\begin{itemize}
\item How much delay did you experience between your actions and their expected outcomes? (Not much -- unendurable)
\item How well did you feel you could manipulate objects in the virtual environment? (Not well -- very well)
\item Did you understand the game? (No -- enjoyable)
\end{itemize}


\subsection{Telekinesis}

In this experiment, we split the virtual table into nine parts, sequentially placing a target in each one of the subsections. Participants were first given up to two minutes to learn the gesture system, then we observed their command choice (including non-gestural physical interaction) for each target position. 

The time elapsed for each choice was recorded, as well as the participant's responses to the following questions on a scale from 1 to 5:

\begin{itemize}
\item How natural did the mechanism for controlling movement through the environment feel? (Not natural -- very natural)
\item How compelling was your sense of objects moving through space? (Not compelling -- very compelling)
\end{itemize}

Following this, the participant's accuracy was tested on one final command, given a target direction. We recorded the participant's response to the following question:
\begin{itemize}
\item Which direction matched your expectation of the command's result the best? (Forward / Backward / Left / Right / None)
\end{itemize}

\subsection{Analysis}

The average length of a \textit{Tic-Tac-Toe} game was one minute and forty seconds for the version using our physical controller, and one minute and five seconds for the purely virtual version. Although the physical version took longer on average, twenty-five percent of all the participants played faster with the physical controller than without it. From this data we concluded that the version using our physical controller was comparable in accessibility to the purely virtual version--that is to say, we did not find a significant difference between the two play-styles. We would postulate that participants were more focused on how to win rather than the method of input. Ideally, our physical controllers are so seamlessly integrated that they are unnoticeable.

\begin{figure}[ht]
  \centering
  \includegraphics[width=0.25\columnwidth]{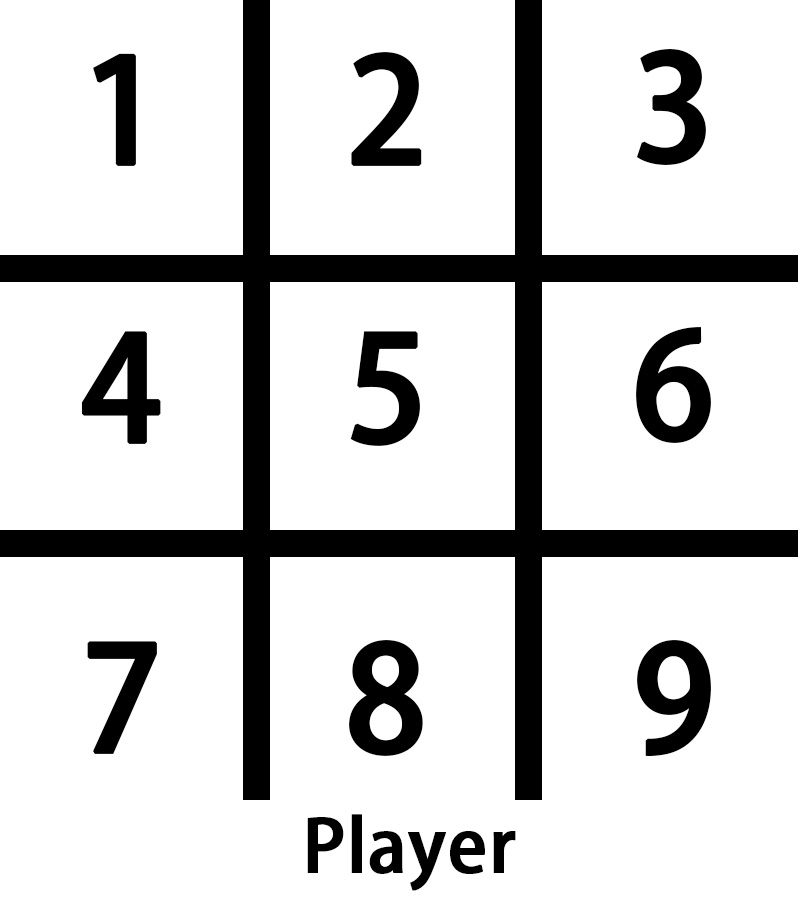}
  \caption{Overview of table tiles with player position}~\label{fig:tabledis}
\end{figure}

\begin{table}
  \centering
  \caption{Amount for the decision on directly grabbing}~\label{tab:grab}
  \begin{tabular}{l c c c c c c c c c}
     \toprule
position on the table &1&2&3&4&5&6&7&8&9\\
    \midrule
amount of grabbing & 3&1&1&4&5&0&6&5&4 \\
     \bottomrule
  \end{tabular}
\end{table}


Table \ref{tab:grab} needed to be explained with Figure \ref{fig:tabledis}, which showed the distribution of tiles. Each tile represents the area where the target would be directed. In the first row of table \ref{tab:grab}, the number representing positions on the table meant the same thing in Figure \ref{fig:tabledis}. The second row showed how many times they chose to grab instead of use gestures to control the target. 

For tiles one, two, and three (far from the player) in Figure \ref{fig:tabledis}, the amount of grabbing is sum of 3,1 and 1, which if 5. We calculated that 5 out of 48 (10.4\%) were moved by grabbing and 43 out of 48 (89.6\%) were moved using gestures. 
Same way for tiles four, five, and six (at a medium distance from the player), 39 out of 48 (81.3\%) were moved using gestures. 
And for the close tiles--seven, eight, and nine--33 out of 48 (68.8\%) were moved using gestures.

The average time elapsed when physically moving the objects was 1.5 seconds (ranging from 0.2 seconds to 5 seconds), whereas the average time elapsed when using gestures was 1.97 seconds (ranging from 0.5 seconds to 3.2 seconds).

From this data, we determined that directly picking up an object is easier than gesturing, especially at close distances, but gesturing is faster if optimized, and is more stable over distance.



While most participants preferred to use gestures for distant targets and physical interaction for proximate targets, one participant mentioned that they preferred gestures over physical interaction because the gestures were more fun. Two other users both chose to use gestures to push proximate targets away before pulling them back again. We concluded from this that the interaction we designed is not only meaningful and useful, but enjoyable as well.




\subsection{Survey}
\begin{table}
  \centering
  \caption{Tic-Tac-Toe Survey}~\label{tab:tictactoe}
  \begin{tabular}{l r}
    {\small\textit{Questions}}
    & {\small \textit{Results}}\\
    \midrule
...different tables? & 2 / 16  \\
...opponent moving naturally? & 16 / 16 \\
...comfortable with the physical version? & 11 / 16\\
...experienced delay? & 2.31 / 5\\
...feel you could manipulate objects? & 4.02 / 5\\
...understand the game? & 4.38 / 5\\
     \bottomrule
  \end{tabular}
\end{table}

\begin{table}
  \centering
  \caption{Telekinesis Survey}~\label{tab:pickup}
  \begin{tabular}{l r}
    {\small\textit{Questions}}
    & {\small \textit{Results}}\\
        \midrule
...felt natural for controlling movement? & 3.5 / 5\\
...compelling sense of objects moving? & 4.0 / 5\\
    \bottomrule
  \end{tabular}
\end{table}

\begin{figure}[h!]
  \centering
  \includegraphics[width=0.7\columnwidth]{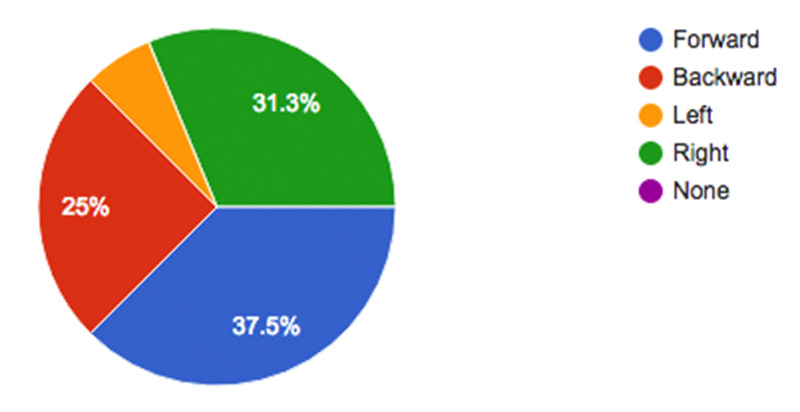}
  \caption{Preference of direction for telekinesis}~\label{fig:FBLR}
\end{figure}

Figure \ref{fig:FBLR} shows the preference of direction for telekinesis. After the tests, participants were asked to chose the easiest direction for them to move. 37.5 percents chose forward because the gesture is easy and the effect is impressive. 31.3 percents chose right rather than left probably because they are all right-handed. We concluded "picking is easier than gesturing for closer" based on the time consumed and their choices. 

\subsubsection{Sense of Reality}
Via Table \ref{tab:tictactoe}, only two participants (12.5\%) felt isolated during the experiment. All users agreed that their opponent moved naturally. From this we conclude that our remote-space representation was acceptable.

\subsubsection{Preference}
Via Table \ref{tab:tictactoe}, eleven of the sixteen participants (68.8\%) chose the physical controller over no controller at all for the "Tic-Tac-Toe" test. Although the value was not as high as we would have liked, we concluded that our physical controller was acceptable.

\subsubsection{Feasibility}
For the question \textit{"How well did you feel you could manipulate objects in the virtual environment"} in Table \ref{tab:tictactoe}, eleven graded a 4, four graded a 5, and one graded a 3 (average 4.02). From this data, we concluded that a robot proxy is acceptable for remote interaction.

In Table \ref{tab:pickup}, for the question, \textit{"How natural did the mechanism for controlling movement through the environment feel?"}, 81\% (average 3.5) of participants felt it was natural enough ($\geq 3$) to control the objects using gestures without actually touching the objects. In addition, the average score for the question, \textit{"How compelling was your sense of objects moving through the space?"} was 4, with 87.5\% grading compelling ($\geq 4$).

\subsubsection{Feeling}
For the question \textit{"How much delay did you experience between your actions and their expected outcomes?"} in Table \ref{tab:tictactoe}, more than 75\% felt it was endurable ($\leq 2$). The average value was 2.31, lower values being better (less perceived latency).

\section{CONCLUSION AND FUTURE WORK}
In this paper, we proposed a new approach for interaction in virtual reality via robotic haptic proxies, specifically targeted towards collaborative experiences, both remote and local. We presented several prototypes utilizing our three mapping definitions, demonstrating that -- using our system -- several virtual objects can be represented physically by one or more robotic proxies, and multiple users can share touch on the same virtual object, or use gestures to command objects without touching them. Our preliminary experiments returned positive results.  In the future we plan to undertake a more comprehensive study focusing on deeper application interactions and expanded hardware capability.

\bibliographystyle{SIGCHI-Reference-Format}
\bibliography{sample}

\end{document}